# Informe Técnico / Technical Report

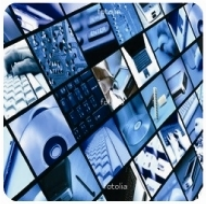 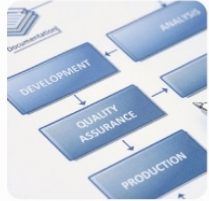 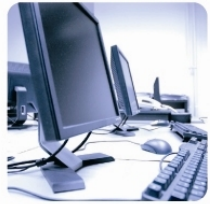 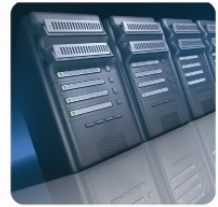 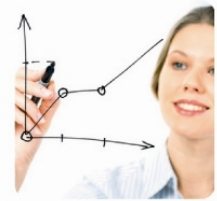

## Communication Analysis modelling techniques

**Sergio España, Arturo González, Óscar Pastor, Marcela Ruiz**

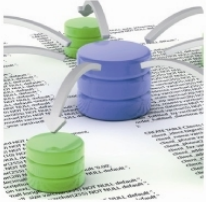 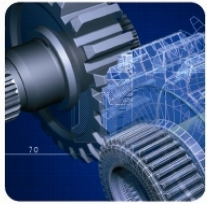 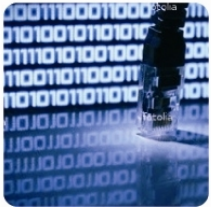 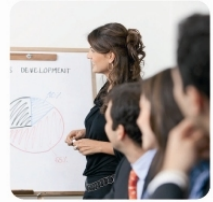 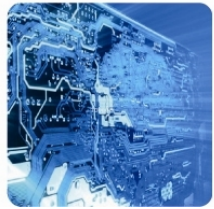

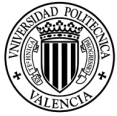 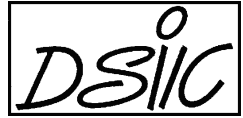 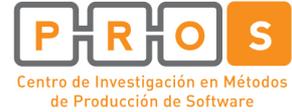

# COMMUNICATION ANALYSIS MODELLING TECHNIQUES

**Authors (in alphabetical order):**
**Sergio España, Arturo González, Óscar Pastor, Marcela Ruiz**

# TABLE OF CONTENTS

# 1     *Scope of this document*

This report describes and illustrates several modelling techniques proposed by Communication Analysis; namely Communicative Event Diagram, Message Structures and Event Specification Templates. The Communicative Event Diagram is a business process modelling technique that adopts a communicational perspective by focusing on communicative interactions when describing the organizational work practice, instead of focusing on physical activities[1]; at this abstraction level, we refer to business activities as communicative events. Message Structures is a technique based on structured text that allows specifying the messages associated to communicative events. Event Specification Templates are a means to organise the requirements concerning a communicative event.

This report can be useful to analysts and business process modellers in general, since, according to our industrial experience, it is possible to apply many Communication Analysis concepts, guidelines and criteria to other business process modelling notations such as BPMN [OMG 2011]. Also, Message Structures can complement business process models created with other notations different than Communicative Event Diagram.

Most of this work has been included in the PhD thesis of Sergio España [España 2011], and some parts of it have been published in academic publications[2].

A platform independent metamodel for Communication Analysis is presented. The metaclasses are not specified in detail; you can find that information in [Ruiz 2011].

Throughout the document, a running example is used. The SuperStationery Co. lab demo is described in full detail in [España, González et al. 2011].

Feel free to contact the authors for additional information or to feedback your experiences with the techniques described herein.

# 2     *Communication Analysis requirements structure*

In Communication Analysis, requirements specifications are mainly organised around the set of communicative interactions that enterprise actors need to perform in order to carry out their tasks. Communicative interactions constitute *what* the enterprise needs in terms of information (the problem); therefore we consider them to be the main requirements to be discovered and described during information system analysis. The rest of requirements qualify communicative interactions by stating aspects, qualities, constraints, etc. of the communication. These solution requirements constitute *how* the system has to be implemented or has to perform. Communication and information requirements are related to *efficacy*; that is, the adequacy of the information supplied to achieve a task. Solution requirements relate to *efficiency*, to the minimisation of operation and usage costs. E.g. usability requirements and response-time constraints are intended to reduce costs. We propose a layered structure for requirements that covers both the problem and the solution spaces. This structure is the backbone that supports Communication Analysis requirements engineering (including the elicitation, the specification and the analysis of requirements).

---

[1] Physical activities such as "A warehouse worker piles up the boxes where the client which may be relevant enough to model them, but always at a lower level of abstraction, using stepwise refinement mechanisms.

[2] A comprehensive list of publications concerning Communication Analysis can be found in http://www.citeulike.org/user/sergioespana/tag/communication_analysis

We define a requirements structure with five levels; each requirement is ascribed to one requirements level. Figure 1 shows the requirements levels, as well as the main activities performed and the artefacts created at each level. The requirements structure offers two dimensions. One dimension is related to the static-dynamic duality (horizontal axis of Figure 1). The other dimension is related to refinement (vertical axis of Figure 1).

In organisational systems, a duality appears among dynamic and static aspects: business interactions are such things because they affect business objects and, in the opposite way, certain objects are considered business objects because organisational interactions deal with them. Communication Analysis facilitates dealing with this duality by offering techniques for interaction analysis (activities 2, 4 and 6) and for the analysis of business objects (activities 3, 5 and 7). This way, following a systemic approach, Communication Analysis allows using stepwise refinements techniques in a twofold perspective. From a dynamic perspective, analysing business processes and obtaining, from the process specification, the business objects structure. From a static perspective, discovering the business objects structure and, then, reasoning the communicative interactions that allow the users to deal with those business objects. This intertwining of both perspectives makes the method flexible and contingent.

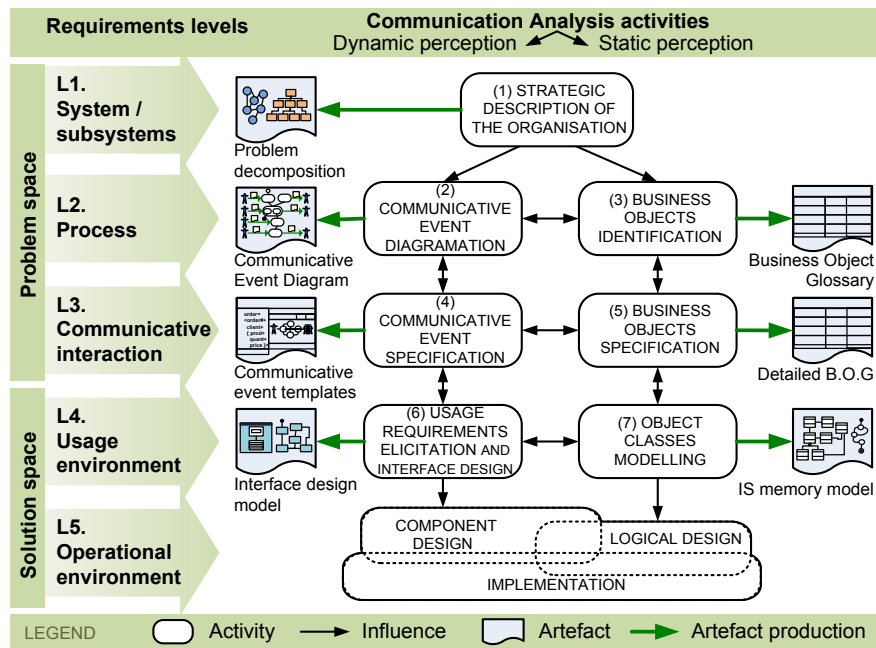

**Figure 1.** The requirements structure and the Communication Analysis workflow

In Communication Analysis, each and every requirement is associated to a specific communicative interaction or to a set of interactions. However, this raises the issue of modularity. How can sub-systems be identified? How many processes should be defined? What is the appropriate granularity of a communicative interaction? When should the analyst stop refining communicative interactions and start designing their support? To answer these questions, Communication Analysis provides guidelines to define requirements models modularly.

In levels L1 and L2, the analyst seeks to refine a complex system into sub-systems and to obtain the repertoire of communicative interactions that the users need for their work practice. From level L2 to level L3 the analyst makes a quantum leap and starts specifying the identified communicative interactions. Another qualitative difference exists between levels L3 and L4, since the analyst (or designer) shifts the focus of the specification from a pragmatic perspective to a semantic and syntactic perspective; the requirements model is added details about how the messages are edited and displayed (i.e. interface design). Also, whereas in L3 the focus was messages (data in motion), L4 focuses on the information system memory (data at rest) intended to ensure the persistence of the communicative interactions. Lastly, the technological architecture is designed and the software application is implemented (this can again involve data in motion in the form of component interaction and network communication, but now the semiotic level is lower).

# 3 Modelling techniques

This report presents several Communication Analysis modelling techniques; namely, the Communicative Event Diagram (see Section 3.1), Message Structures (see Section 3.2) and Event Specification Templates (see Section 3.3). These techniques are the result of evolving and adapting previously existing techniques to fit the communicational perspective that Communication Analysis imposes on information systems modelling. For instance, the Communicative Event Diagram is a revision of business process modelling techniques (e.g. UML Activity Diagrams) so as to take into account communicative interactions, and Message Structures is a revision of the BNF notation and Structured Analysis data dictionaries. The notations have been tuned up according to our experience throughout the years, leaving out any superfluous details. Nevertheless, other notations can be used instead of the ones that we propose herein, as long as they are conveniently adapted according to the philosophy of the method; that is, as long as the Communication Analysis criteria and guidelines are preserved and the notation is added any lacking modelling primitives.

We place a special emphasis in the methodological guidelines. For instance, the unity criteria for business process modelling (see Section 3.1.2) guide business process model modularity. The guidelines for using Message Structures explain their distinct application during analysis than during design (see Section 3.2.3).

## 3.1 Communicative Event Diagram

The Communicative Event Diagram is intended to describe business processes from a communicational perspective.

### 3.1.1 Grammatical constructs

A *communicative event* is the organisational action that is triggered as a result of a given change in the world (i.e. in the subject system), intended to account for that change by gathering information about it. In each communicative event, someone conveys a message containing new meaningful information to the information system (this message is specified by means of a message structure in level L3). Communication Analysis offers guidelines to allow identifying communicative events (communicative event unity criteria), also enabling modularity. This way, a communicative event can be seen as an ingoing communicative interaction that fulfils the communicative event unity criteria.

Each communicative event is represented as a rounded rectangle and is given an identifier and a descriptive name (see Figure 2). The *identifier* serves for traceability purposes and it is usually a code composed of a mnemonic (related to the organisational system or the business process to which the event is ascribed) and a number (e.g. S*ALE* 1, where S*ALE* is the acronym of the *Sales management process* and 1 is just a number that should be unique within this particular process).

Methodological guideline for naming communicative events:

❖ With regard to the *name*, we recommend to consistently use either an external nomination (primary actor + action + object + qualifier; e.g. "A client places an order") or an internal nomination (interface actor + action + object + qualifier; e.g. "The salesman receives an order"). For instance, in the SuperStationery case we have opted for an external nomination.

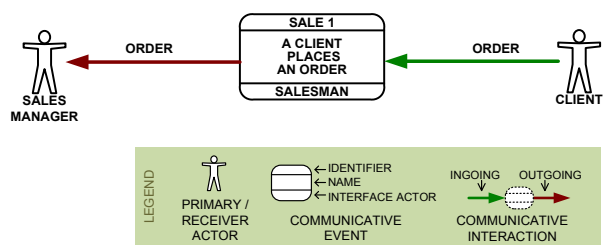

**Figure 2.** Example of communicative event from the SuperStationery case

Organisational roles are types of (human or non-human) actors that participate in the organisational work practice. For instance, members of the enterprise (John, Mary, Frank, Joseph), customers (Phil, Nathalie) and suppliers (Phil, Christine) are organisational actors; salesperson, customer and supplier are organisational roles. Several organisational actors can play the same organisational role (e.g. both Mary and Frank are salespersons). Also, the same actor can play different organisational roles (e.g. Phil is both a customer and a supplier). Organisational roles are assigned a set of rights and obligations; for instance, responsibilities (the duties that organisational actors belonging to that role ought to perform), the type of information they are entitled to know or manage, the set of other organisational actors they are in charge of, etc. For instance, "salespersons attend clients when they place orders; to do so…"

For each event, the organisational roles that play the different communicative roles are identified. Communication Analysis distinguishes several communicative roles:

The *primary role* of a communicative event is responsible for communicating the new meaningful information to the information system (e.g. a client that formulates a request); they are the owners of such information. It is usual that the primary role also triggers the communicative interaction (i.e. they initiate the interaction by establishing contact with the organisational system; e.g. a client comes by and places an order), but it is not always the case (e.g. a sales representative can visit a client and convince him to place an order). Therefore, for each communicative event, the primary role needs to be identified.

Organisational roles playing the primary role are modelled by means of a sticky figure. Since primary actors provide the conveyed input information, they are modelled as senders of ingoing communicative interactions. For instance, the client is the primary actor of event S<small>ALE</small> 1.

The *receiver role*s model one or several organisational roles that need to be communicated the occurrence of an event. The organisational roles playing the receiver role are also modelled by means of a sticky figure. Since these correspond to the actors that need to be informed of the occurrence on an event, they are modelled as receivers of outgoing communicative interactions. In order to truly understand the meaning of messages in organisations, it is necessary to analyse these actors. For instance, in S<small>ALE</small> 1 the sales manager is informed of the order placement.

The *interface role*s model which organisational roles are in charge of physically editing the message in the information system interface (e.g. by typing the data in a softweare screen, or by writing in a paper form).

The organisational roles playing the interface role are specified textually at the bottom of the communicative event rounded rectangle. For instance, the salesman is the interface actor in S<small>ALE</small> 1, because they fill the order form according to the client request. Although the concept of interface actor belongs to the requirements level L4. Usage environment, according to our experience it is often useful to identify interface actors when discovering the business processes. Also, organisational actors often feel more motivated to participate in the analysis when they see their organisational role represented in the business process models.

With regards to support actors, these are not specified in the communicative event diagram, but in the event specification template (see Section 3.3).

Methodological guidelines for communicative roles:

- ❖ For each communicative event, the organisational roles playing the primary role need to be identified (i.e. it is mandatory to have at a primary role). Primary actors are the primary source of the information conveyed to the information system. For each communicative event ask the stakeholders who is providing the new information being coveyed. Clarify that it is not who types the data but who provides it.
- ❖ It is not strictly necessary to model the interface roles. Also, they may change if a technological reengineering takes place (e.g. if a web application is provided we could place the responsibility of typing the orders to the customers themselves).
- ❖ It is not strictly necessary to model the receiver roles. However, doing it provides a useful overview of how information flows within the organisation. For a communicative event ask the stakeholders who should be aware of the occurrence of such event (e.g. who should know that a client has placed an order?). Receiver roles are often primary roles of subsequent communicative events, but not always.

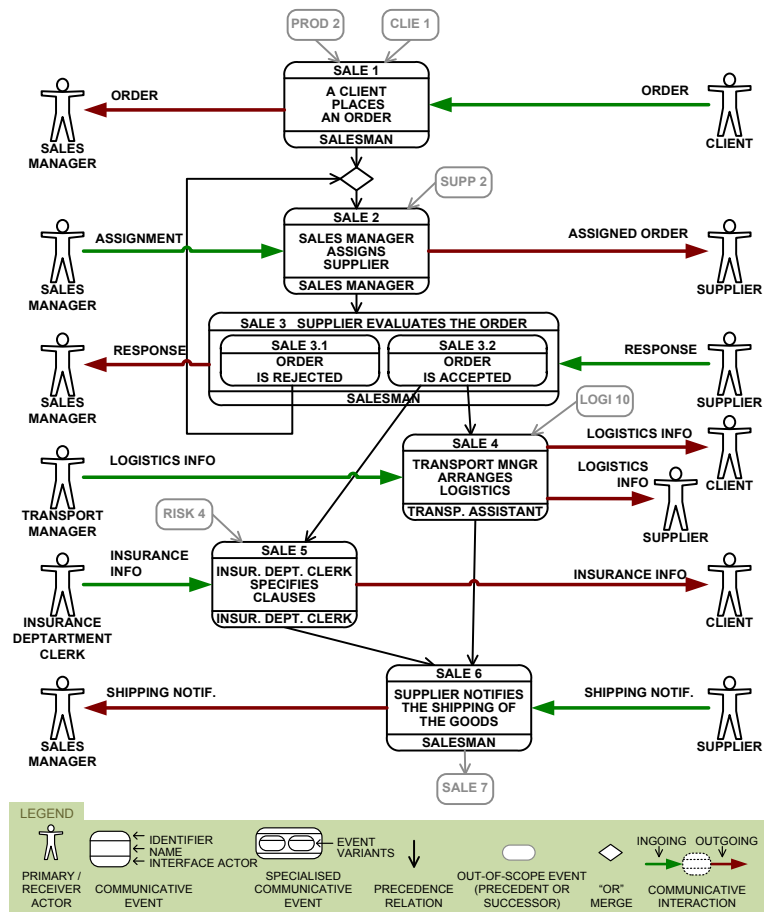

**Figure 3.** Communicative event diagram of part of the *Sales management* business process of SuperStationery case

Do not confuse the primary actor with the interface actor: the primary actor provides the information whereas the interface actor edits the message in an interface (e.g. writes in a paper form or types in a software screen). The primary role and interface role may be played by the same or distinct actor, who may belong to the organisation or its environment.

For instance:

– A client has a desire and phones the company so as to formulate an order (because she is providing the information, she is playing the primary role); the salesman who attended the client in the phone enters the order in the system (because of this message editing, he is playing the interface role).
– The client could have used the electronic form the company has in its website to make the order (this way, a person who is not member of the organisation plays the two roles,).
– In any case, later in time, the sales manager herself assigns the order to a supplier (in this case, a member of the organisation plays the two roles).

Secondary notation guidelines for communicative roles:

❖ The sticky figures representing primary and receiver actors should be laid out in a way that communicative interaction arrows can be routed horizontally, either at the left or at the right of their corresponding communicative event rounded rectangles.
❖ As long as possible, sticky figures should be aligned in vertical columns, trying to minimise the number of columns, and trying to place the roles belonging to the organisational system in one side and the roles belonging to its environment in the other side.

- ❖ If it is perceived to improve diagram comprehensibility, for each organisational role a different vertical column can be created.

Communication between partners often takes place in the form of a dialog (a set of message transfers going back and forth), but in Communication Analysis dialogs are abstracted and referred to as communicative interactions. The communicative interactions associated to communicative events are represented by means of thick lines with a long filled arrowhead pointing at the receiver of the associated message. Depending on the main direction of communication, the following types of communicative interactions can be distinguished:

- *Ingoing communicative interactions* primarily feed the information system memory with new meaningful information. For instance, the placement of an order by a client, the reporting of a fault in a packer machine by an operator, the decision of the director of the Department of Risk Management to accept an investment. These interactions are often (but not always) supported by business forms. *Ingoing communicative interactions* depart from a primary actor sticky figure and arrive at the communicative event rounded rectangle.
- *Outgoing communicative interactions* primarily consult information system memory. These interactions are often materialised as business indicators, listings and printouts. For instance, the list of clients with a debt that is greater than 6000€, a cash flow chart, a receipt, a payroll list. *Outgoing communicative interactions* depart from a communicative event rounded rectangle and arrive at a receiver actor sticky figure.

Communicative interactions are given a name by labelling them. The name should coincide with the name of the associated message structure (unless there is a good reason not to do so) since this is believed to improve cognitive traceability[3].

Secondary notation guidelines for communicative interactions:

- ❖ Communicative interactions should be as straight and horizontal as possible.
- ❖ The colour of ingoing and outgoing communicative interactions should be different. Green and red are the author's choice, but different colours can be chosen. In any case, colour is just used as a visual cue and it is the direction of the arrow that indicates the type.

Communicative events are the main type of nodes in a Communicative Event Diagram. Other types of nodes are logical nodes (i.e. and-forks, and-joins, or-branches and or-merges), start nodes and end nodes.

With regards to the behaviour of the organisation, organisational norms usually define the set of states in which the occurrence of each communicative event is admissible. An intensional definition of such set is usually referred to as the precondition of the communicative event. Currently, there is no formal language for specifying event precondition formulas; they are typically stated textually. However, preconditions often establish temporal relationships among communicative events and, in Communication Analysis, we are interested in precedence relations among communicative events (for which a graphical notation is offered).

A *precedence relation* between two communicative events A and B indicates that event A must necessarily occur before B occurs. This can as well be expressed as "A is a direct precedent (communicative event) of B" or "B is a direct successor (communicative event) of A.

Precedence relations are modelled by means of thin lines with a short open arrowhead, departing from a source node and arriving at a target node. Communicative event A is said to be a precedent of communicative event B (and B a successor of A) if there is a precedence relation that departs from A and arrives at B, or if there is a sequence of precedence relations with only logical nodes in between.

---

[3] Message structures are later specified in detail.

A precedence relation that departs from a communicative event, e.g. A, and arrives at another communicative event, e.g. B, is a *loopback* if A is a precedent of B, and the depth of A is smaller than the depth of B (the depth of a communicative event being the length of the minimum path between the initial node and the event).

Secondary notation guidelines for precedence relations:

- ❖ Precedence relations should be drawn as vertical as possible, being successor events placed lower than precedent events.
- ❖ In general, straight lines are preferred to curved or angled lines. However, angled lines are preferred in the case of loopbacks.

Note that communicative interactions are placed in the horizontal axis, whereas the vertical axis is reserved for precedence relations.

Logical nodes have the following modelling primitives:

- *Or*. The or-merge indicates that only one of the incoming precedence relations needs to hold. Note that the or-branch is implicit and corresponds to event specialization. The or-merge is represented by a diamond shape containing the propositional connective symbol ∨.
- *And*. The and-fork and the and-join are implicitly represented by two or more precedence relations departing from or arriving to a communicative event, respectively; however, they can be explicitly drawn if needed to express complex logical expressions. The and-fork and the and-join are represented by a diamond shape containing the propositional connective symbol ∧. Only one precedence arrow can arrive at an and-fork, but several arrows can depart from it. Several arrows can arrive at an and-merge, but only one arrow can depart from it.

Secondary notation for logical nodes:

- ❖ When or-merges are used in a diagram but no and-forks are used, then the propositional connective symbol can be omitted, as long as the legend consistent with this decision (see Figure 3).
- ❖ The symbols + and x could be used instead of ∨ and ∧, as long as the legend reflects this decision.

Two special nodes can be used at the analyst's convenience.

- A *start node* symbol indicates that the communicative events to which it is connected do not have any precedent events; that is, they are globally initiatory events. The start node is represented by a filled black circle.
- An *end node* symbol represents that the communicative events to which it is connected do not have any successor events; that is, they are locally terminatory events. The end node is represented by two concentric circles, the external one filled in white and the internal one filled in black.

*Globally initiatory events* often result in the creation of a representation of a business object that does not refer to other business objects; for instance, CLIE 1 creates a new client record, SUPP 2 creates a new supplier record, and PROD 2 creates the product records of the company catalogue.

We refer as *locally initiatory events* to communicative events that, having precedent events, result in the creation of a representation of a business object that refers to other business objects; for instance, SALE 1 creates a representation of a new client order (which refers to existing clients and existing products).

Secondary notation for start and end nodes:

- ❖ Start- and end-node symbols are often omitted for the sake of diagrammatic economy. Every communicative event not having any precedent event is implicitly connected to the start node. Every communicative event not having any successor event is implicitly connected to the end node.

A specialised communicative event indicates that different alternatives in the path through the business process are possible[4]. Specialisation can be induced externally, either by structural or domain variants in the message structure that is associated to the event, or by a decision explicitly made by the primary actor. Specialisation can also be induced internally, as a result of previous events. We refer as *event variant* to each alternative behaviour within a specialised communicative event. We advice using an encapsulated notation for specialisation (i.e. to graphically encapsulate all the event variants within the specialised communicative event) to reinforce the notion of unity. Each event variant has a corresponding *specialisation condition*, which is a well-formed formula that can refer to one or several fields of the message structure (in case the specialisation is externally induced).

Specialised events are represented by rounded rectangles that contain one smaller rounded rectangle per each of the event variants, as is SALE 3 (see Figure 3). Several levels of specialisation are possible; that is, an event variant can, in turn, be specialised (see Figure 4). Both the specialised event and the event variants can be target and source of precedence relations. For instance, for ALIE 5.2.2 to occur, ALIE 4 and SPAC 4 need to have occurred first; after the occurrence, ALIE 8, ALIE 10 or ALIE 11 can occur. In every case, the captain will take note of the alien's decision.

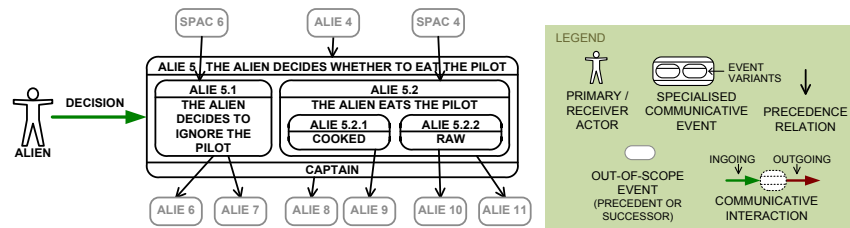

**Figure 4.** A specialised communicative event with two specialisation levels

Secondary notation for event specialisation:

❖ It is convenient to align horizontally all the event-variant rounded rectangles of the same level.

Methodological guidelines for event specialisation:

❖ Communicative events are specialised whenever each event variant leads to a different temporal path (i.e. distinct precedence relations depart from each variant).

❖ It must be avoided specialising an event as a result of different communication channels, since the message remains the same (e.g. a publishing house can order a report in person or by telephone).

The business processes of an organisational system can be specified in a single communicative event diagram or in several ones, depending on the size and complexity of the resulting diagrams. When the model denotation has been partitioned into several diagrams, there will probably exist precedence relations among communicative events of different diagrams. Although the preconditions of each communicative event are also described in event specification templates, it is important to make these precedence relations explicit in the diagrams. To do so, out-of-scope precedent events and *out-of-scope successor events* can be included in a communicative event diagram, representing them by means of a small grey rounded rectangle only containing the event identifier (they simply act as a reference). For instance, Figure 3 specifies part of the *Sales management* business process; that is, the process contains other communicative events concerning the reception of the shipping and the invoicing, but they have been omitted to keep the diagram simple. Note that SALE 7 is left out of the scope of the diagram but it has been represented with the proposed symbol to indicate that the business process continues after the notification of the shipping departure. Also, all the direct precedents of the communicative events of the diagram have been included, even if they belong to different business processes; namely, *PROD 2*, *CLIE 1*, *SUPP 2*, *LOGI 10* and *RISK 4*. Clie 1 is included in the *Client management* business process (see Figure 5). See another example of diagram partitioning in Section 3.1.3.

---

[4] This type of specialisation is somehow related to business process variability, although further investigation is needed to clarify the similarities and the differences between both notions.

Secondary notation for diagram partition:

❖ Precedence relation arrows departing or arriving at out-of-scope events can be either black or grey, but the colour should be consistent within each diagram. This gives the arrows more or less relevance but does not change their meaning.

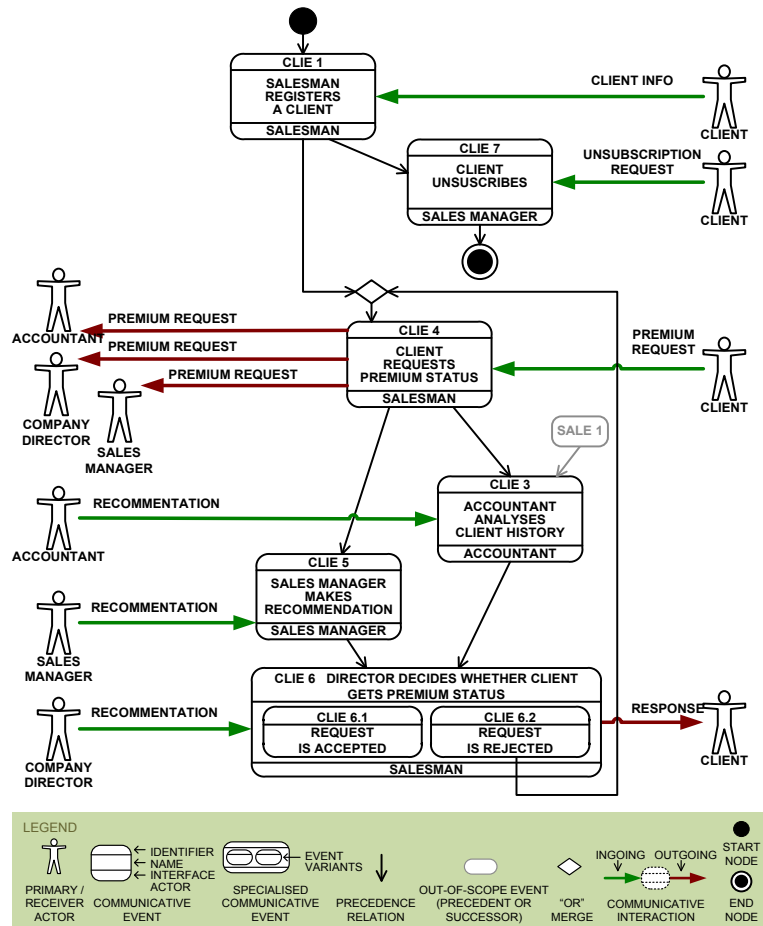

**Figure 5.** *Client management* business process of the SuperStationery case.

Methodological guidelines for diagram partition:

❖ The out-of-scope direct successors can be omitted but it is important to include all the out-of-scope direct precedents. The reason is that precedence relations are part of the precondition of the successor event. This way, it is important to represent in the diagram in Figure 3 that, for *SALE 2* to occur, *SUPP 2* must have occurred first. On the contrary, although it is convenient, it is not indispensable to represent in this diagram that after *SALE 6*, *SALE 7* can occur. It will be important, however, to represent this precedence relation in the second half of the model; that is, when *SALE 7* is included in a diagram, then *SALE 6* should also be included even in it is done by means of the out-of-scope symbol. Also note that, in Figure 5, it is indicated that communicative event *CLIE 3* has event *SALE 1* as precedence but it has not been considered important to specify that *SALE 1* has *CLIE 1* as precedence (*SALE 1* is the out-of-scope direct successor of *CLIE 1*).

- There exist guidelines that suggest partitioning a big business process diagrams. For instance, 7PMG recommends decomposing the diagram when it has more than 50 elements [Mendling, Reijers et al. 2010]. These guidelines should be followed, unless there is a strong reason not to do so.

Last but not least, it is strongly advised to add a legend to each and every communicative event diagram. This way, if the image is ever displayed out of its original context or presented to an audience that is unaware of the notation, the symbols would be easier to decipher.

- The legend should include each and every symbol used in the diagram. If a symbol does not appear (e.g. no specialised communicative events appear in Figure 2, so its symbol can be omitted in the legend).

For the Communicative Event Diagram, we recommend the notation described above. However, many aspects of the notation can be changed if needed (e.g. the shape for communicative events can be a circle instead of a rounded rectangle), as long as the changes are reflected in the legend.

### 3.1.2 Guidelines for business process model modularity

Business process modelling methods provide notations that often include a means for activity refinement. However, most of the times the methods lack precise guidelines for model modularity. As a result, business process models end up mixing activities of different abstraction levels in the same diagram (we have tried to represent this in Figure 6.a). We claim that appropriate unity criteria aids analysts in creating modular models. We consider that a business process model has an appropriate modularity with respect to a given set of unity criteria when it fulfils the criteria; this may include having a structure of modelling layers and using stepwise refinement: each activity belongs to a specific modelling layer and can be refined in a subsequent layer if needed (we have tried to represent this in Figure 6.b).

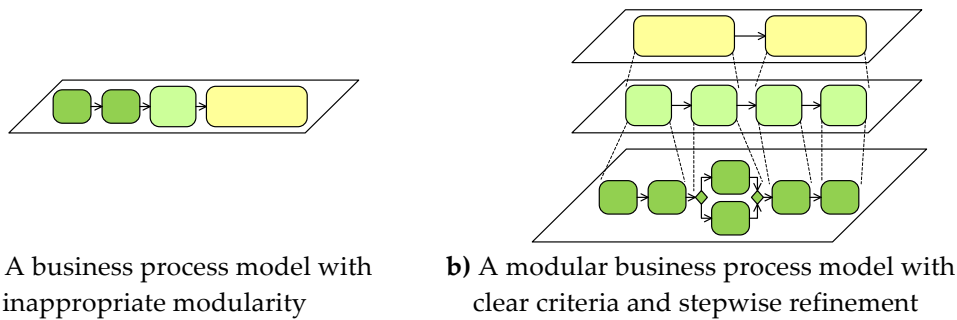

**a)** A business process model with inappropriate modularity

**b)** A modular business process model with clear criteria and stepwise refinement

**Figure 6.** Different communication levels in information systems

The unity criteria that are proposed in this section are based on principles from Systems Theory and Communication Theory. In order to observe, control and influence the subject system, the organisational system performs business processes. Business processes demand much organisational communication; for this reason, organisations rely on information systems, which are often computerised for the sake of efficiency.

With regards to Communication Theory, we build upon two well-founded models; namely, Shannon's model of a general communication system [Shannon 1948] and Jakobson's model of the communicative act [Jakobson 1990].

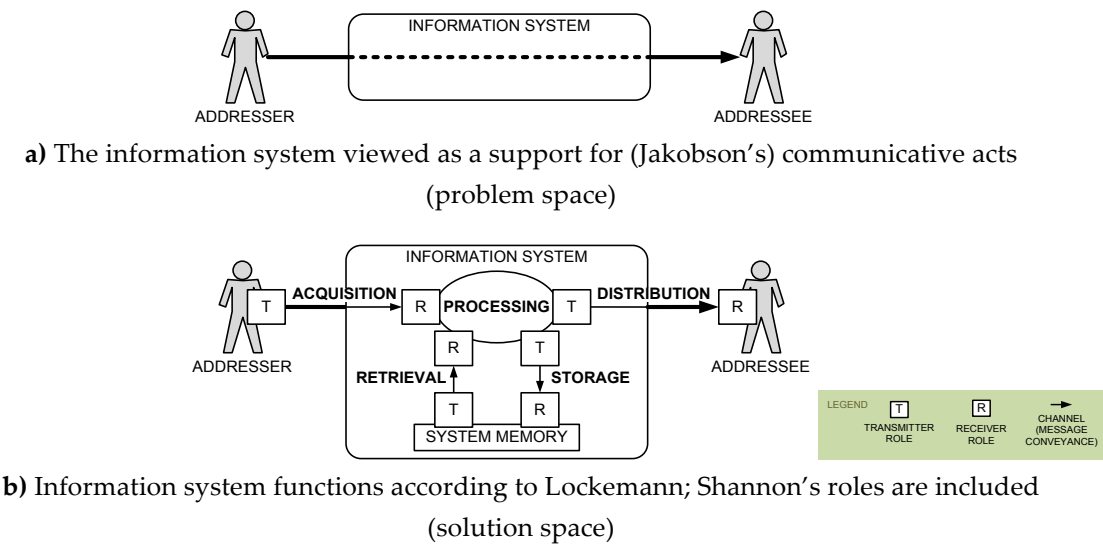

**a)** The information system viewed as a support for (Jakobson's) communicative acts

(problem space)

**b)** Information system functions according to Lockemann; Shannon's roles are included

(solution space)

**Figure 7.** Different communication levels in information systems

From a communicational perspective, an information system can be considered a communication channel among actors of the organizational system or its environment (the subject system). This vision is depicted in Figure 7.a. A communicative act takes place, via the information system channel, between an addresser and an addressee[5].

According to Lockemann and Mayr, information system functions are the acquisition, retrieval, processing, storage and distribution of information [Lockemann and Mayr 1986]. Figure 7.b refines Figure 7.a in order to reflect these functions. A similar model is included in the FRISCO report [Falkenberg, Hesse et al. 1998], and in the extension to that framework presented in [España 2011, Chapter 3]. At this level of communicational abstraction, many message conveyances appear. Shannon's model is applicable since, in each message conveyance, a transmitter (T) and a receiver (R) intervene.

Needless to say, actual work practise can involve more complex communication procedures (such as re-encodings, translations, and chains of message conveyances that do not add new information). E.g. information acquisition may involve a client formulating a request to the clerk, and the clerk editing that message in order to convey it to the information system.

The existence of several communicational levels and complex communication procedures adds complexity to business process modelling, since business process encapsulation is different according to each level. The unity criteria defined next, clarify these levels (and, thus, facilitates business process modelling) by providing guidance for encapsulation.

We propose analysing organisations and their information systems following a systemic and communicational approach. From a systemic point of view, the focus is put on external interactions, in order to determine organisational behaviour. From the communicational point of view, unity criteria based on communication functions are proposed. Table 1 summarises modularity as applied to business process modelling.

The intention of modularity in business process modelling is to define capsules that correspond with external interactions. The content of such capsules (i.e. internal system composition) determines organisational reaction to external interactions. Business process modularisation allows distinguishing between problem space (i.e. organisational needs that are independent of any particular implementation) and solution space (i.e. the support that allows addressing organisational needs efficiently).

---

[5] Only the addresser, the channel and the addressee are depicted, the remaining elements from Jakobson's model are kept implicit

Table 1. Modularity in business process modelling

|  | **Business Process Modelling** |
|---|---|
| **Intention** | To distinguish and achieve independence between problem space and solution space (a.k.a. the 'what' and the 'how', the communicational need and the support to communication), in communication-oriented business process models. |
| **Encapsulation** | Unity criteria are based on the notion of complete communication; that is, trigger unity, communication unity, and reaction unity (see Table 2). Unitive conceptions (and, by means of metonymy, their representations) are referred to as communicative events[6]. |
| **Information hiding** | When focusing on communicative events at the problem space, internal composition and internal reaction are disregarded. Business process models at this level specify organisational behaviour. |

We refer as communicative event to a set of actions related to information, which are carried out in a complete and uninterrupted way, on the occasion of an external stimulus. Similar definitions can be found in [ISO 1987; Yourdon 1989; Falkenberg, Hesse et al. 1998]. In other words, an event happens in the environment of the information system, someone communicates this occurrence to the information system and a series of synchronous activities is triggered. Table 2 defines unity criteria that facilitate the identification of communicative events and the specification of business processes. Each criterion corresponds to one function of communication, as defined by Jakobson.

Table 2. Unity criteria to identify and encapsulate communicative events

| **Criterion**          (*Communication function*)<br>Definition | Type of communicative event | |
|---|---|---|
|  | Logical | Physical |
| **Trigger unity**                                    (*Phatic function*)<br>Trigger responsibility is external.<br>This means that event occurs as a response to an external interaction and, therefore, some actor triggers it. This (primary) actor is the one that provides the information that is conveyed in the event. | ✓ | ✓ |
| **Communication unity**                        (*Referential function*)<br>The input message must constitute a non-empty, complete unit.<br>Firstly, this means that each and every event involves providing new meaningful information. Thus, an interaction needs to provide new facts in order to be considered an event. Input messages are representations of something that happens in the subject system.<br>Secondly, this means that the message must be complete. Thus, it should be avoided to define several events for communicating parts of a message that only makes sense for the organisational system as a whole. | ✓ |  |

---

[6] Although we refer as communicative events to the unitive conceptions related to business process modelling that are obtained by using the proposed unity criteria and their representations, actually it is the criteria which are important, not the name. This way, as long as unity criteria are applied, capsules can be referred to as tasks, activities, processes, use cases, etc.

| | | | |
|---|---|---|---|
| **Reaction unity**                              (*Connative function*)<br>The event is a composition of synchronous activities; thus, these activities can communicate the information they need from each other. Events are asynchronous among each other; thus, events need a shared information system memory to communicate. | | | ✓ |

These unity criteria allow marking out a frontier between problem space and solution space. They allow differentiating logical events and physical events[7]:

- *Logical events* correspond to Jakobson's communicative acts. They are ascribed to the problem space. Trigger unity and communication unity allow encapsulating this type of events[8]. Therefore, they are also referred to as communicative events. Logical events provide knowledge and their encapsulation disregards communicational support agents such as transmitters and receivers, only addressers and addressees are taken into account (see Figure 7.a). These events are of great importance[9].

- *Physical events* correspond to Shannon's message conveyances. They are ascribed to the solution space. Trigger unity and reaction unity allow encapsulating this type of events. Physical events are activities that provide concrete support to communication; thus, they depend on support agents such as transmitters and receivers (see Figure 7.b). These events can be changed without altering communicational content.

Starting with the idea that an occurrence takes place in the subject system, communication unity is closer to the external stimulus, and reaction unity is closer to information system reaction (communication comes before reaction).

Given a set of communicational needs of the organisational system, the set of logical events that satisfy those needs can be established objectively. On the other hand, physical events that support the logical ones depend on technological constraints, organisational responsibilities, budgetary constraints, etc.

For instance, physical events can depend on organisational responsibilities. The synchronism mentioned in the reaction unity criterion is conditioned to processing actors. Processing actors are those actors that are involved in a logical event but do not provide new information (only the primary actor does provide new information). Processing actors are typically in charge of carrying out Shannon-related activities (e.g. encoding, transmission, decoding) and Lockemann-related activities (e.g. acquisition, retrieval, processing, storage, distribution, see Figure 7.b). This way, an organisational system can decide to assign the responsibility of processing tasks to one or several workers. In case several workers are assigned the responsibility of processing tasks, this implies a change in the work environment; that is, the business process flows from one work environment to another. Then, several physical events appear, at least one for each work environment.

Communication Analysis business process unity criteria can be used to create business process models from scratch, but also to re-structure existing models.

*Illustrative example of the application of unity criteria*

The following is a case description that partly based on a real case; it is inspired by the work practice of a governmental institution (the name of the institution is fictitious) as performed some decades ago (processes have been improved over the years). The case illustrates the application of the unity criteria to unravel complex business processes by to separating the wheat from the chaff; that is, differentiating logical and physical events.

---

[7] Physical vs. logical distinction in organisational system modelling already appears in DeMarco's work [1979].

[8] A logical event can fulfil the three criteria; this would result in a much optimised information system. Unfortunately, this is not always the case.

[9] Logical events (communicative events) subsume physical events and they are the main focus during business process modelling in Communication Analysis.

Valencia is a Spanish region that has an important agricultural sector (e.g. oranges). In case of hail or hard frost, the Valencia Institution of Agriculture and Farming (VIAF, a governmental institution) receives help requests from farmers. Each farmer goes to their corresponding town council and manually fills in a form with personal data, data about damaged cultivated plots of land, and the amount of money corresponding to damages in each plot. Farmers hand over the form to a council clerk. After the closing date, each council sends the received forms to VIAF. Then, a Technical Department clerk transfers the forms to a company that digitises data (message encoding is outsourced). The company returns digitised data and original forms to VIAF Technical Department, where digitised data is inspected and detected errors are corrected. Subsequently, help requests are recorded in the database. Then an agricultural technician visits the farmer's plots in order to assess the damages, take some photographs and handwrite some notes. Later, the agricultural technician issues an expert report. In view of the help request and the expert report, the Secretary of the Crisis Management Department resolves whether to grant economic aid and, if so, determines the amount. A Technical Department clerk sends a letter that informs the farmer about the resolution, and a copy of the resolution is sent by fax to the corresponding town council.

Figure 8 shows the above-mentioned process in terms of logical events, whereas Figure 9 further refines logical event *VIAF1* in order to depict its associated physical events[10]. Logical events *VIAF2* and *VIAF3* also have a similar model of their corresponding physical events, but these are not included herein for the sake of brevity.

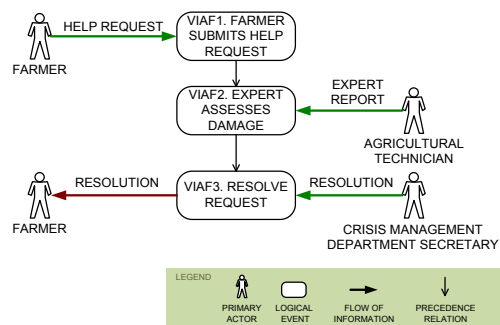

**Figure 8.** Logical events of the VIAF business process model for help requests

---

[10] Purposely, a loose notation has been used in the models. They are only intended to clarify the notions of logical and physical events. We believe that these notions are applicable to many business process modelling notations.

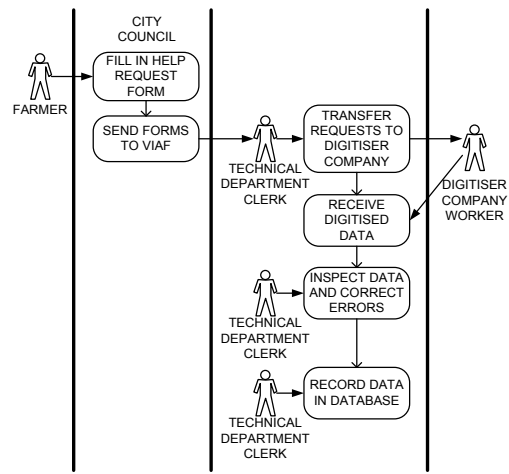

**Figure 9.** Physical events corresponding to logical event *VIAF1*. *Farmer submits help request*

### 3.1.3 Diagram partitioning

The Communicative Event Diagram modelling technique allows interrelating two or more diagrams by using out-of-scope symbols and precedence relations. The out-of-scope symbol has been presented in Section 3.1.1, and guidelines for its usage have been provided. Whenever a communicative event from one diagram has a precedent event that belongs to a different diagram, this relation has to be specified (it is important to have readily at sight all the precedent events because precedence relations are part of an event precondition). Also, whenever a communicative event has a successor event that belongs to a different diagram, then this relation can be specified (although in this case it is not so relevant, it helps conceiving the big picture of the organisational work practice).

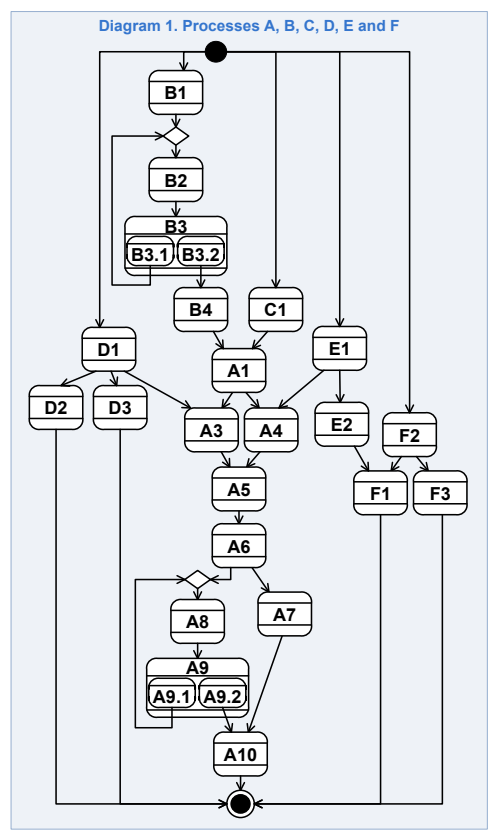

**Figure 10**. A diagram presenting a complete view of the processes of the whole business

By means of defining several interrelated communicative event diagrams, the analyst can partition the description of organisational work practice. The benefit is that cognitive complexity is reduced because the fragmented models are more manageable; this way the modelling technique becomes scalable. The trade-off lies in the fact that the out-of-scope precedence relations have to be kept consistent.

For instance, Figure 10 shows an illustrative (meaningless) example of a company's work practice. The work practice has been divided into several business processes, but all of them are depicted in the same communicative event diagram. The business process model can be partitioned into several diagrams, using different partitioning criteria, as shown in Figure 11. For instance, each business process has been modelled in a separate diagram. Also, business process A has been considered two big and has been further partitioned in two parts. All the diagrams in Figure 10 and Figure 11 are just views of the complete business process model (which, if supported by a CASE tool, can be stored in a repository).

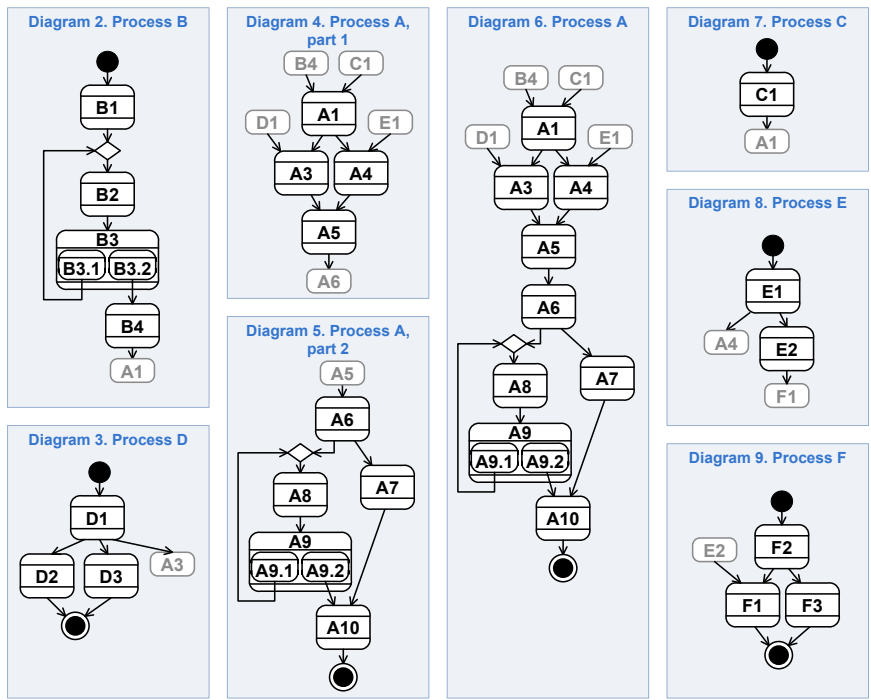

**Figure 11.** Several diagrams presenting partial views of the whole business process model

## 3.2 Message Structures

Message Structures is a specification technique that allows describing, by means of structured text, the message that is associated to a communicative interaction. Although message structures can be used to specify outgoing communicative interactions, we focus on the specification of ingoing communicative interactions, given their analytical interest. Table 3 presents an example[11] of the usage in analysis time; the ORDER message structure specifies a communicative interaction by which a client places an order[12].

Table 3. Example of a message structure in analysis time

| FIELD | OP | DOMAIN | EXAMPLE VALUE |
|---|---|---|---|
| ORDER = | | | |
| < Order number + | g | number | 10352 |
| Request date + | i | date | 31-08-2009 |
| Payment type + | i | text | Cash |
| Client + | i | Client | 56746163-R, John Papiro Jr. |
| DESTINATIONS = | | | |
| { DESTINATION = | | | |
| < Address + | i | Client address | Blvd. Blue mountain, 35-14A, 2363 Toontown |
| Person in charge + | i | text | Brayden Hitchcock |
| LINES = | | | |
| { LINE = | | | |
| < Product + | i | Product | ST39455, Rounded scissors (cebra) box-100 |
| Price + | i | money | 25,40 € |
| Quantity > | i | number | 35 |
| } | | | |
| > | | | |
| } | | | |
| > | | | |

### 3.2.1 Grammatical constructs

The syntax of Message Structures can be described in terms of the following grammatical constructs.

We refer as *substructure* to an element that is part of a message structure. This way, LINE, Client and Payment type are substructures that are part of ORDER. There exist two classes of substructures: fields and complex substructures. We refer as *initial substructure* to the substructure that constitutes the first level of a message structure. For instance, ORDER=<Order number + Request date + Payment type + Client + DESTINATIONS >.

- *Field.* It is a basic informational element of the message; that which is not composed of other elements. There exist two types of fields.
  o *Data field.* It is a field that represents a piece of data with a basic domain[13]. For instance, Payment type is a data field that belongs to the message structure ORDER.
  o *Reference field.* It is a field whose domain is a type of business objects. E.g., Client references a client that is already known by the information system.
- *Complex substructure.* It is any substructure that has an internal composition. There exist three types of complex substructures.

---

[11] This particular font, the colours and the capitalisation are a non-prescriptive convention that is intended to facilitate message structure comprehension. Feel free to configure these aspects.

[12] Most of the examples in this paper are taken from a requirements model that can be found in http://hci.dsic.upv.es/ca/SuperStationery-TR-v2.0.pdf

[13] Basic domains (e.g. numbers, text) are discussed below.

- o *Aggregation substructure.* It specifies a composition of several substructures in a way that they remain grouped as a whole. It is represented by angle brackets `< >`. For instance, LINE=<Product+Price+Quantity> specifies that an order line consists of information about a product, its price and the quantity that the client requests.
- o *Iteration substructure.* It specifies a set or repetition of the substructures it contains. It is represented by curly brackets `{ }`. For instance, an order can have several destinations and, for each destination, a set of order lines is defined. Both DESTINATIONS and LINES are iteration substructures. LINES={LINE=<Product+Price+Quantity>}
- o Specialisation substructure. It specifies one or more variants; that is, structural alternatives[14]. There is no example of specialisation substructure in Table 3; the message structure in Table 4 specifies that the assignment made by an student can be of type THEORY, in which case the fields Subject and Title characterise the work, or it can be of type PRACTICE, in which case the fields Programming language and Functionality characterise the work.

Table 4. Fragment of a message structure that includes specialisation

| FIELD | OP | DOMAIN |
|---|---|---|
| <   ... |  |  |
|    Type of assignment + | i | [theo\|prac] |
|    TYPE = |  |  |
|    [ THEORY = |  |  |
|       < Subject + | i | Subject |
|         Title > | i | text |
|    \| PRACTICE = |  |  |
|       < Programming language + | i | Language |
|         Functionality > | i | text |
|    ] +   ... |  |  |
| > |  |  |

For greater disambiguation, Table 5 presents the grammatical constructs of Message Structures using the Extended Backus-Naur Form notation (EBNF) [ISO/IEC 1996].

In practice, the syntax is more flexible: the names of complex substructures can be omitted, an iteration substructure also aggregates its own content (there is an implicit aggregation substructure), and each variant of a specialisation substructure also has an implicit aggregation. This 'syntactic sugar' allows adapting the notation to project contingencies and it facilitates the usage of the technique.

Table 5. EBNF grammar of Message Structures[15]

```
message structure
= structure name, '=', initial substructure;
initial substructure
= aggregation substructure | iteration substructure;
aggregation substructure
= '<', substructure list, '>';
iteration substructure
= '{', substructure list, '}';
specialisation substructure
= '[', substructure list,{ '|', substructure list },']';
substructure list
= substructure, { '+', substructure };
complex substructure
= aggregation substructure | iteration substructure
| specialisation substructure;
substructure
= substructure name, '=', complex substructure | field;
```

---

[14] It is more frequent to use specialisation with two or more variants. The usage with one variant represents the optionality of that variant; that is, a message might or might not include the variant.

[15] The elements `structure name`, `substructure name` and `field` are character strings.

## 3.2.2 Field specification

To characterise a field, the following properties can be specified:
- *Name*. Each field must have a significant name (e.g. Request date).
- *Acquisition operation*. It specifies the origin of the information that the field represents.
  - *Input* i. The information of the field is provided by the primary actor.
  - *Generation* g. The information system can automatically generate the information of the field.
  - *Derivation* d. The information of the field is already known by the information system and, therefore, it can be derived from its memory; that is, it was previously communicated in a preceding communicative event. This operation can have an associated derivation formula.
- *Domain*. It specifies the type of information the field contains.
- *Example*. An example of a value for the field, provided by the organisation.
- *Description*. An explanation that helps the reader to understand the field meaning.
- *Label*. A brief text that describes the field when shown in a graphical interface.
- *Link with memory*. It specifies the correspondence between the field and a database table column or a class diagram attribute.
- *Compulsoriness*. It specifies whether the field necessarily takes value or not (i.e. whether the field is mandatory or not). It is also possible to specify that the field is not compulsory by using a one-variant (e.g. [a]).
- *Initialisation*. The value that the field is given by default can be specified by means of a function or a derivation formula.
- *Visibility*. It specifies whether the field is visible in a graphical user interface form.

It is recommended to lay the fields out vertically ad to specify the field properties horizontally (by means of columns). For reasons of space, the description of the fields can be done in a separate table. Message Structures can be extended with other field properties that a method designer or an analyst deem appropriate. However, as discussed below, not all properties are convenient at analysis time.

## 3.2.3 Usages of Message Structures

**Table 6.** Applicability of field properties to development stage

| | | Name | Acquisition operation | | | Domain | Example | Description | Label | Link with memory | Compulsoriness | Initialisation | Visibility |
|---|---|---|---|---|---|---|---|---|---|---|---|---|---|
| | | | i | g | d | | | | | | | | |
| Analysis | | ++ | ++ | ++ | -- | ++ | ++ | ++ | -- | -- | -- | -- | -- |
| Design | Memory | ++ | ++ | ++ | ++ | ++ | ++ | ++ | - | ++ | + | - | - |
| | Interface | ++ | ++ | ++ | ++ | ++ | ++ | ++ | ++ | ++ | ++ | ++ | + |

++ highly recommended   + recommended   - not recommended   -- discouraged

Message Structures can be applied for different purposes (from software development to adaptive maintenance) and in different stages of the software development life cycle (e.g. analysis, design). Depending on whether they are used in analysis or design time, syntactic and pragmatic differences have to be taken into account. Table 6 presents recommendations on the usage of field properties, depending on the development stage in which Message Structures are used.

**Creation and usage of Message Structures in analysis time**

In analysis time, Message Structures allow specifying in detail the communicative interactions that take place in the organisational work practice. This way, they offer a communicational perspective for business process modelling and they act as requirements for the information system. In the context of Communication Analysis, the new meaningful information that is conveyed to the information system in each communicative event is specified by means of a message structure.

In the following, we enumerate some sources of information and techniques for acquiring information and analysing the messages exchanged with the information system.

*Organisational actors* play an important role in information systems analysis, since they know organisational work practice first-hand. The analyst will employ elicitation techniques such as interviews or JAD sessions [August 1991]. It is crucial to ask the proper questions so as to define which information is conveyed in each communicative event, as well as to distinguish new information from derived information.

*Business forms* are a technological support for communicative interactions and, therefore, they are a major source for analysis. In this sense, the user interface screens from pre-existing software are equivalent. Forms can be used for entering information (input forms), for presenting data (output forms), or for both purposes. In analysis time, input forms allow to identify communicative events that convey new information to the information system. See [España 2011] for an explanation on how to analyse input forms to find out the communicative interactions they support and to specify their corresponding message structures.

If the organisation has *previous business process specifications* or quality procedures, then this documentation can also be used as input for the analysis.

**Usage of Message Structures in design time**

In design time, Message Structures allow establishing the traceability between analysis documentation, the specification of the information system memory (e.g. by means of a class diagram or a relational database schema), and the specification of the user interface. Moreover, it is possible to define techniques for deriving the memory if the information system from requirements models, as well as techniques for systematically reasoning the interface design.

The summarised procedure for the derivation of the information system memory is as follows[16]. First the communicative events are sorted according to their temporal precedence. Then the message structure of each event is processed in order to obtain a class diagram view. This way, the complete class diagram is iteratively created by integrating the class diagram views that correspond to all the communicative events. Figure 12 (right-hand side) shows the derivation of the class diagram view that corresponds to a communicative event in which a client places an order. A more detailed and bigger example is available online[12].

The summarised procedure to reason the user interface is as follows. First the interface style manual. Then the editing environments are identified (i.e. sets of forms or interface screens that support a set of editorially-compatible communicative events. Next, the message structures are fragmented (e.g. normalising them in first normal form) and the fragments are assigned to abstract interface structures (e.g. registry, set of registries). The abstract interface structures are encapsulated in forms. Each form is specified in detail, establishing the possible interaction and the editing facilities (filters, order criteria). The behaviour of the interface can be specified by means of trigger tables. Lastly, additional listings and printouts are specified. Figure 12 (left-hand side) shows how the information system interface is designed in terms of abstract interface patterns. Methodological guidelines are described in detail in [España 2005].

---

[16] We describe the derivation of class diagrams because this derivation technique is part of ongoing research. An analogous argumentation can be made for relational schemas.

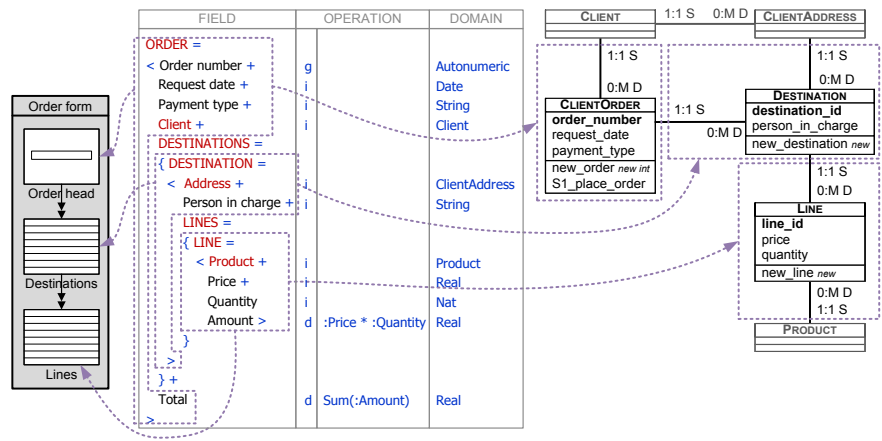

**Figure 12.** Derivation of a class diagram view and the interface design of a communicative event in which an order is placed

In design time, it is usual to specify derived fields (e.g. the total amount of each order line Amount d (:Price * :Quantity)). Other properties that specify aspects of design are also specified in this stage (e.g. the specification of the data field Request date could also include a formula that defines its initialisation value: today()).

## 3.3 Event Specification Templates

Communicative events that appear in the Communicative Event Diagram need to be described in detail. Requirements associated to an event can be structured by means of an event specification template. The template is composed by a header and three categories of requirements: contact, message and reaction requirements. These categories are related to Jakobson's phatic, referential and connative communication functions, respectively [Jakobson 1990].

The *header* contains general information about the communicative event; that is, the event identifier, its name, a narrative description and, optionally, an explanatory diagram.

- *Event identifier*. Event identification needs to be kept consistent throughout the entire analysis and design specification in order to enhance requirements traceability. This way, the event identifier needs no to coincide with the one in the communicative event diagram. It normally consists of the acronym of the business process to which the event belongs and a sequential number.
- *Event name*. The name of the communicative event should clearly state the change in the subject system that is being reported in the event. See 3.1.1 for naming guidelines.
- *Narrative description*. Since requirements specifications are meant, first of all, to facilitate problem understanding, a narrative description of the event is strongly advised.
- *Explanatory diagram*. Also, whenever the event is complex, an explanatory diagram illustrating its associated flow of tasks shall be included. In such diagram, we recommend to focus only in physical communicative events; that is, actions related to information acquisition, re-encoding, and distribution that fulfil the trigger unity and the reaction criteria. In Section 3.1.2 the unity criteria are explained and Figure 9 presents an explanatory diagram depicting physical events.

*Contact requirements* are related to the conditions that are necessary in order to establish communication. For instance, the primary actor, possible communication channels (e.g. fax, email, in person), availability and temporal constraints (e.g. office hours for order reception), authentication requirements (e.g. in Spain, bureaucratic proceedings often require showing an identity card).

- *Primary actor*. The organisational role that is responsible for communicating with the information system to report a change in the subject.
- *Support actors*. The organisational roles that participate in message transfers but do not provide new information; that is, the actors responsible for physical communicative events.
- *Interface actors*. The organisational roles that are in charge of editing input messages in the form and codes required by the information system. In a paper-supported information system, the interface actor is the one that fills paper business forms; in a computerised information system, the interface actor is the one that interacts with the user interface of the software application.
- *Availability requirements* and constraints that refer to the degree to which the information system is in a position to engage in the ingoing communicative interaction.
- *Medium requirements* that refer to the technology (this includes paper-based forms) that supports the ingoing communicative interaction. In case scanned business forms or screenshots of a previous software application are available, those that apply to the communicative event (e.g. those that support the ingoing communicative interaction) can be included in this section. In case they are catalogues in a different document or repository, then a reference to them can be included.
- *Accreditation requirements* that refer to the protocols that the organisational system prescribes for each actor participating in the ingoing communicative interaction (i.e. informally, how to know that actors are who they say they are).

- *Verification requirements* that refer to a) ensuring that the provided documentation (if any) is not fraudulent, and b) confirming the influx of physical elements associated to the ingoing communicative interaction (e.g. stock coming into a warehouse).

*Message requirements* specify the message conveyed by the primary actor to the information system in a communicative event and related constraints (e.g. reliability: certifying that a diploma provided by a student is not fraudulent). With regard to the message, both metalinguistic aspects (e.g. message field structure, compulsoriness of fields) and linguistic aspects (e.g. field domains, example values) need to be specified. We propose Message Structures to specify the message (Table 7 presents a summary of the grammatical constructs, see Section 3.2 for more details), but the analysts can choose an equivalent notation. In any case, the following aspects are the most important and should be specified whatever notation is chosen.

- *Metalinguistic aspects*. They refer to the structure of the message, its editing and display.
  - *Structure of the message* that is conveyed to the information system; that is, its composition in terms of complex substructures and message fields.

Table 7. Summary of the grammatical constructs of Message Structures

| Message Structure grammatical constructs |
|---|
| **Aggregation**     A = < a + b + c > <br> A is composed of fields a and b and c. |
| **Alternative**     A = [ a \| b \| c ] <br> A is either composed of field a or b or c, (only one of them). |
| **Iteration**     A = { B } <br> A is composed of several substructures of type B. |

  - *Acquisition operation*. Indicating whether the data contained by a field is new information for the organisational system or simply a recall of previously-reported information (this includes derived information such as total amounts) is important in order to analyse whether the event is actually a communicative event or it should be discarded for not providing new meaningful information.
  - *Description of the fields*. A description should be provided for the sake of comprehensibility.
- *Linguistic aspects*. They refer to the content of the message and its meaning.
  - *Domains* of the message fields. During analysis, an orientation of the field content should be given (preferably not a programming-language data type).
  - *Example value.* One or several realistic values that the field can contain clarify its meaning; they should be provided by the organisational actors involved in the analysis (i.e. representative users).
  - *Derivation* of field values. Some fields (and even some complex substructures) are derived from already-known information. Specifying such derivation can be done textually or by means of formulas.
- *Message constraints*. Such as constraints over the structure of the message, over the domains of the message fields, etc.[17]

---

[17] Sometimes the message (the structure or the field domains) needs to be particularised to organisational roles or actors depending on their characteristics (e.g. their duties, the information that concerns them, etc.). For instance, privacy constraints on the composition of the message determine which fields an actor can/cannot view (e.g. in a given company, only salespersons can view the fields 'discount' and 'commission' and not their assistants), constraints over the domain restrict the information domain to which an actor has access (e.g. a salesman can only view orders placed on its own region).

A communicative event cannot be fully understood until the structure of its ingoing message is defined in detail. Specifying with precision an event message structure forces and helps analysts and users to appropriately mark the boundary of an event and its meaning for the organisation.

*Reaction requirements* describe how the information system reacts to the communicative event occurrence (i.e. to the conveyed message). Typically, the information system processes and stores the new information (updating the system memory), extracts all the necessary conclusions that can be inferred from new knowledge, and makes new knowledge and conclusions available to the corresponding actors (distributing the information to other actors so that they can act accordingly). Therefore, this category of requirements includes the treatment or processing of the information and the outgoing communicative interactions being generated by the event, among other requirements.

- *Data model view* related to the communicative event; it is the part of the memory of the information system that the communicative event contributes to build. This is an aspect of design that can be specified by means of linking each message field with tables and columns (in case the memory is being specified by means of a relational schema), with classes and attributes (in case object orientation is chosen), etc. In the context of the integration of Communication Analysis and the OO-Method, the data model view is actually a class diagram view (see [González, España et al. 2011]).

- *Treatments* that define what changes occur in the information system as a result of the communicative event (e.g. what processing takes place, what information is stored). It involves defining how the information acquisition is related to the data model. In some cases is suffices to indicate that the information is stored. In other cases, complex processing needs to be done and, thus, an algorithm needs to be described with more or less level of detail (textually, with pseudocode, etc).

- *Linked behaviours* refer to how the occurrence of a communicative event affects future occurrences of events. For instance, the information provided in this communicative event can condition future behaviours of the system. This includes business rules or complex conditions (e.g. decision tables) that determine future reactions depending on the values provided in the current communicative event.

- *Linked communications* specify to whom the occurrence of the communicative event must be communicated; that is, which organisational roles (or specific organisational actors) need to know about the event occurrence and its associated information so as to take further actions (e.g. make decisions). In many cases, during analysis, it may suffice to state the actor who needs to be reported the occurrence; in such cases, linked communications can simply be expressed as outgoing communicative interactions in the communicative event diagram. If further details about the communication need to be specified, it can be done in this section (e.g. scanned output form, sketch).

In Figure 13, the event specification template corresponding to event SALE 1 of the SuperStationery case is provided as an example.

### SALE 1. A CLIENT PLACES AN ORDER

## 1  General information

**Goals**

The objective of the organisation is to attend the clients when they request goods.

From the point of view of the information system, the objective of this event is to record the order that the client places, and to let the Sales Manager know that a new order has arrived.

**Description**

Most clients call the Sales Department, where they are attended by a salesman. Then the client requests one or several products that are to be sent to one or many destinations. The salesman takes note of the order. Other clients place orders by email or by fax.

## 2  Contact requirements

**Actor responsibilities**

- **Primary actor**: Client
- **Communication channel**: In person, by phone, by fax
- **Interface actor**: Salesman

**Temporal requirements**

- **Occurrence temporal constraints**: Only working days during reception hours (09:00-18:00)
- **Frequency of occurrence**: 500 orders per week

**Business forms**

**ORDER**

Order number: 10352  
Payment type: ☒ Cash ☐ Credit ☐ Cheque  
Request date: 31-08-2009  
Planned delivery date: 05-09-2009

**Client**  
VAT number: 56746163-R  
Name: John Papiro Jr.  
Telephone: 030 81 48 31

**Supplier**  
Code: OFFIRAP  
Name: Office Rapid Ltd.  
Address: Brandenburgen street, 46, 2983 Millhaven

Destination: Blvd. Blue mountain, 35-14A, 2363 Toontown  
Person in charge: Brayden Hitchcock

| # | Code | Product name | Price | Q | Amount |
|---|------|-------------|-------|---|--------|
| 1 | ST39455 | Rounded scissors (cebra) box-100 | 25,40 € | 35 | 889,00 € |
| 2 | ST65399 | Staples cooper 26-22 blister 500 | 5,60 € | 60 | 336,00 € |
| 3 | CA479-9 | Stereofoam cups box-50 (pack 120) | 18,75 € | 10 | 187,50 € |
|   |        |             |       |   | 1412,50 € |

Destination: Greenhouse street, 23, 2989 Millhaven  
Person in charge: Luke Padbury

| # | Code | Product name | Price | Q | Amount |
|---|------|-------------|-------|---|--------|
| 1 | ST65399 | Staples cooper 26-22 blister 500 | 5,60 € | 30 | 444,50 € |
| 2 | CA746-3 | Sugar lumps 1kg | 2,30 € | 3 | 6,90 € |
|   |        |             |       |   | 451,40 € |

Total | 1863,90 €

Form 1. Example of an order form

Some parts of the form are not yet filled in this event.

## 3 Message requirements

**Message structure**

| FIELD | OP | DOMAIN | EXAMPLE VALUE |
|---|---|---|---|
| ORDER = | | | |
| < Order number + | g | number | 10352 |
| Request date + | i | date | 31-08-2009 |
| Payment type + | i | text | Cash |
| Client + | i | Client | 56746163-R, John Papiro Jr. |
| DESTINATIONS = | | | |
| { DESTINATION = | | | |
| < Address + | i | Client address | Blvd. Blue mountain, 35-14A, 2363 Toontown |
| Person in charge + | i | text | Brayden Hitchcock |
| LINES = | | | |
| { LINE = | | | |
| < Product + | i | Product | ST39455, Rounded scissors (cebra) box-100 |
| Price + | i | money | 25,40 € |
| Quantity > | i | number | 35 |
| } | | | |
| > | | | |
| } | | | |
| > | | | |

| Field | Description |
|---|---|
| Order number | A sequential number that identifies the order. |
| Request date | The date in which the client places the order. |
| Payment type | Information about the payment type. Its value is normally either Cash, Credit or Cheque, but the salesman can freely indicate any other information here. |
| Client | The client that places the order. |
| Address | A client destination at which the products have to be delivered. |
| Person in charge | The name of the person that will receive the order at the destination. From one order to another, this person can be different. |
| Product | A product that is requested by the client. |
| Price | The price of the requested product. |
| Quantity | The amount of items of the product that the client requests at a specific destination. |

**Structural constraints**

One order can have many destinations.

One destination can have many lines.

One order is places by exactly one client

**Contextual constraints**

Orders are identified by Order number.

The product price in the line takes its value from the current price of the product in the catalogue.

## 4 Reaction requirements

**Treatments**
The order is recorded.

**Linked communications**
The Sales Manager is informed of the order placement.

**Figure 13.** Example of an event specification template (both pages)

# 4 Communication Analysis platform-independent metamodel

As part of the method definition, we propose a metamodel that contains many of the constructs that are defined by the method specifications in previous sections. The metamodel is depicted in Figure 14. The metamodel includes the metaclasses intended for modelling the dynamic view of the information system. It mainly covers the three first requirements levels: L1 concerned with organisational modelling and problem decomposition, L2 concerned with business process modelling from a communicational perspective, and L3 concerned with communicative event specification.

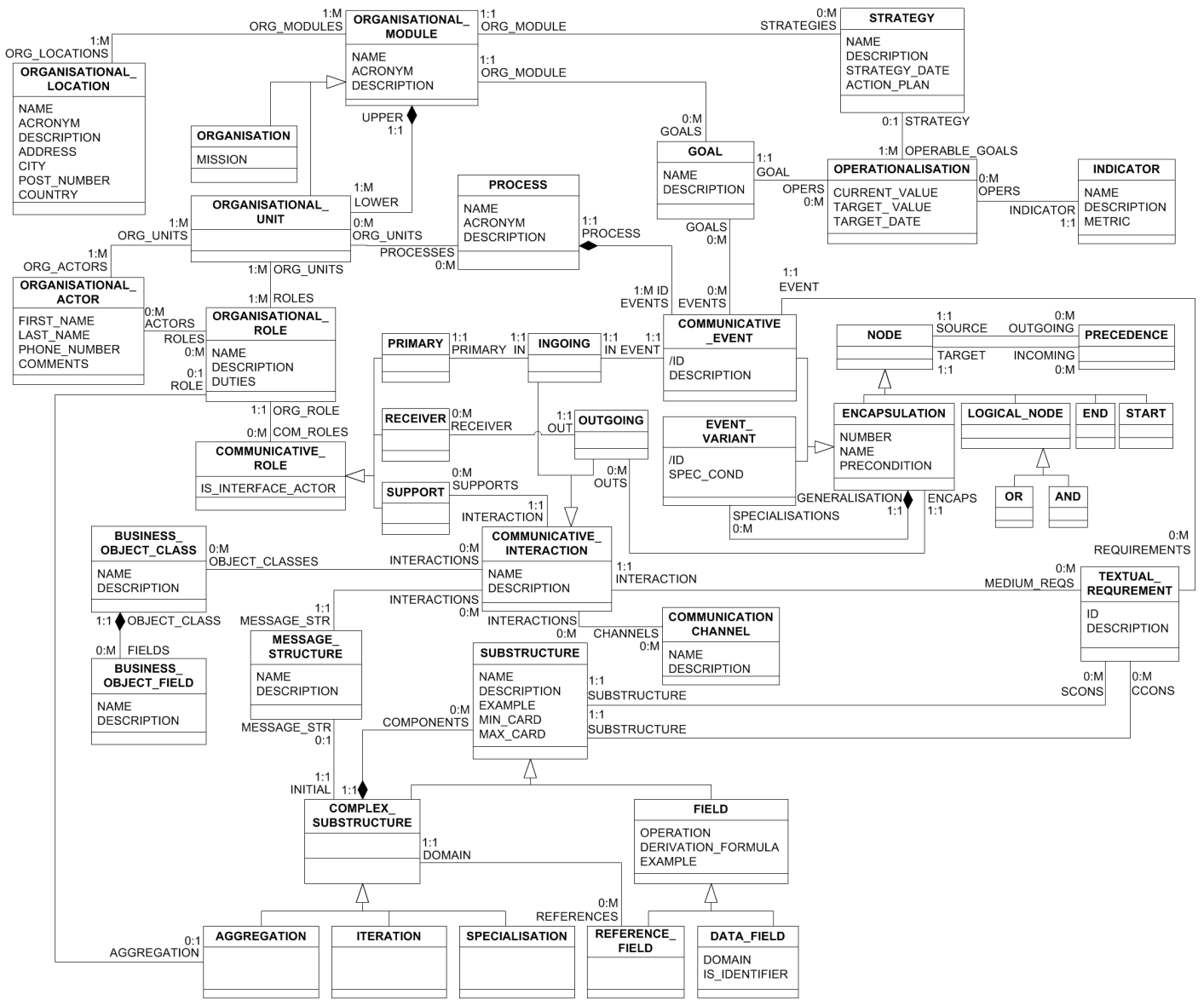

**Figure 14.** Communication Analysis platform-independent metamodel

We have omitted some metaclasses intended for modelling the static view of the information system because this aspect is expected to be covered by metaclasses of the OO-Method after the integration of both methods (e.g. metaclasses related to normalised business object class information definitions; that is, those metaclasses intended for modelling the computerised information system memory).

Take into account that the taxonomy of textual requirements is just provided as an exemplification of our vision that every requirement in a requirements model should be categorised properly. For implementation purposes, this specialisation hierarchy could be replaced by a more flexible way of supporting the taxonomy.

Table 8 presents two formulas corresponding to the derived attributes that calculate the identifier of communicative events and event variants. This way, events of the *Sales management* process will have identifiers of the form S<small>ALE</small> 1, S<small>ALE</small> 2, S<small>ALE</small> 3, etc. Also, the variants of the specialised event S<small>ALE</small> 3 have a compound identifier of the form S<small>ALE</small> 3.1 and S<small>ALE</small> 3.2.

**Table 8.** Derivation formulas for the Communication Analysis PIM metamodel

| Derivation Formulas | |
|---|---|
| F1 | C<small>OMMUNICATIVE</small>_E<small>VENT</small>.Id = concat( G<small>ENERALISATION</small>.Id, IntToStr( Number ) ) |
| F2 | E<small>VENT</small>_V<small>ARIANT</small>.Id = concat( P<small>ROCESS</small>.Acronym, IntToStr( Number ) ) |

The following constraints restrict what can be expressed using the metamodel in Figure 14. These rules are part of the grammars of Communication Analysis modelling languages.

**Table 9.** Constraints for the PIM metamodel of Communication Analysis

| Constraints | |
|---|---|
| C1 | A start node must not have incoming precedence relations. |
| C2 | An end node must not have outgoing precedence relations. |
| C3 | An and node must have two or more incoming precedence relations and only one outgoing precedence relation. Take into account that the and node represents the and-join, since the and-split is implicit (see Section 3.1.1). |
| C4 | An or node must have only one incoming precedence relation and two or more outgoing precedence relations. Take into account that the or node represents the or-merge, since the or-branch is implicit in communicative event specialisation (see Section 3.1.1). |
| C5 | Within each business process, each communicative event must have a distinct number. |
| C6 | Within each specialised communicative event, each event variant must have a distinct number. |
| C7 | A formula is either a specialisation condition, an initialisation formula or a derivation formula, but not several of these at the same time (this means that an instance of F<small>ORMULA</small> can only have a link via one of the relationships). |
| C8 | The initial substructure of a message structure cannot be a specialisation substructure. Thus, given an instance of C<small>OMMUNICATIVE</small>_E<small>VENT</small>, it cannot be linked via Message_Structure to a C<small>OMPLEX</small>_S<small>TRUCTURE</small> of type S<small>PECIALISATION</small>. |
| **C9** | The minimum cardinality of a message structure field (an instance of Field) is 0 or 1, and the maximum cardinality of a field is 1. |
| **C10** | The minimum and the maximum cardinality of an aggregation substructure (an instance of Substructure) is 1. |
| **C11** | The minimum and the maximum cardinality of a specialisation substructure (an instance of Specialisation) is 1. |

# 5 Summary

This report has presented a detailed specification of Communication Analysis modelling techniques and their associated metamodel, with the intention to pave the way for its integration with a model-driven development method; namely, the OO-Method. Communication Analysis provides a framework for eliciting, analysing and modelling information systems from a communicational perspective. It proposes a requirements structure with five levels (L1 to L5). Each level has been explained, its underlying concepts have been defined on top of the FRISCO 1.1 ontology (presented in [España 2011]), and some examples of its application have been provided. L1. System/Subsystems deals with the structure and the goals of the organisational system. L2. Process deals with discovering and modelling business process models in terms of communicative events. L3. Communicative Interaction specifies in detail each communicative event, including its associated message. L4. Usage Environment specifies the interface and the memory of the information system. L5. Operational Environment enters architectural design. With regards to the application of Communication Analysis for model-driven development, both L4 and L5 are out of the scope, because the OO-Method will cover these levels once both methods are integrated (see [España 2011; España, González et al. 2011; España, Ruiz et al. 2011; González, España et al. 2011]).

The report has also presented with full detail the modelling techniques that Communication Analysis proposes for requirements specification:

- Communicative Event Diagram, a business process modelling technique that adopts a communicational perspective and facilitates the development of an information system that will support those business processes (see Section 3.1). The most important feature of the technique is a set of unity criteria that guides model modularity (see Section 3.1.2), since it can be applied to other modelling notations as well.
- Message Structures, a modelling technique for the specification of messages communicated with (and within) the organisation (see Section 3.2).
- Event Specification Templates, a textual requirements specification technique intended to support the description of communicative events (see Section 3.3).
- In order to facilitate the integration of Communication Analysis into a model-driven development framework, a metamodel of the method has been provided (see Section 4). The metamodel presented in this chapter is platform independent, in the sense that it does not yet regard any CASE-tool development platform and it purely focuses on clarifying the constructs of the modelling techniques (a metamodel that is specific for the Eclipse Modeling Framework is presented in [Ruiz, España et al. 2010]).